\documentclass{article}
\usepackage{graphicx}
\usepackage{cite}

\begin{document}

\title{Excitation of zonal flow by the modulational instability in
electron temperature gradient driven turbulence}

\maketitle

\author{Yu A Zaliznyak$^1$, A I Yakimenko$^{1,2}$ and V M Lashkin$^1$}

\textit{$^1$ Institute for Nuclear
Research, Nauki ave., 47, Kiev 03680, Ukraine \\
$^2$ Department of Physics, Taras Shevchenko National University,
prospekt Glushkova 2, Kiev 03680, Ukraine}

\begin{abstract}
The generation of large-scale zonal flows by small-scale
electrostatic drift waves in electron temperature gradient(ETG)
driven turbulence model is considered. The generation mechanism is
based on the modulational instability of a finite amplitude
monochromatic drift wave. The threshold and growth rate of the
instability as well as the optimal spatial scale of zonal flow are
obtained.
\end{abstract}



\section{Introduction}

It is now an established fact that zonal flows (ZFs) (i.e.,
azimuthally symmetric modes that depend only on the radial
coordinate) play a crucial role in regulating the nonlinear
evolution of drift-wave instabilities in tokamaks, and
consequently, the level of turbulent transport
\cite{DiamondRev1,DiamondRev2}. In particular, it is known that
triggering of the L-H transition in tokamaks is related to the
emergence of ZFs in the poloidal direction which suppresses the
fluctuations and build up a barrier to the turbulent transport. It
is widely thought that zonal flows are generated by the
modulational instability of the turbulent spectrum of
electrostatic drift-wave perturbations \cite{Smol1,Smol2}. Under
this, two possible regimes can be indicated: (a) when the spectrum
is broad, integration over all wavenumbers yields a resonant
instability; (b) when the drift wave spectrum is narrow, one can
consider the instability of a monochromatic drift wave and the
instability is of the modulational type. In the latter case,
standard schemes of the four-wave interaction can be applied
\cite{Chen,Manfredi,Zonca}.

Ion temperature gradient (ITG) modes and trapped electron modes
are generally regarded as the main candidates to explain the
anomalous transport, but, recently, there has been the growth of
interest in electron temperature gradient driven turbulence,
produced by the electron temperature gradient driven modes
\cite{HollandDiamond,Gurcan1,Gurcan2,Gurcan3}. This is related to
the fact that the study of interaction of ETG modes with
large-scale motions like ZFs or streamers (radially elongated
vortex-type structures) is important for understanding the
electron thermal transport within an internal transport barriers,
when ITG turbulence is suppressed by $\mathbf{E}\times\mathbf{B}$
shear flow \cite{Jenko}. Note, that there is an important
difference between the ITG and ETG models: in the ETG model we
have the Boltzmann ion response for both waves and zonal flows,
while the electron response to zonal flow perturbations is
hydrodynamic in the ITG model \cite{DiamondRev1,Weiland}.

In the present paper we consider the excitation of ZFs by a finite
amplitude monochromatic drift wave in the framework of ETG
turbulence model. The corresponding ETG mode is assumed to be
stable, i. e. we consider the region below the marginal stability
boundary. We derive a set of coupled equations describing the
nonlinear interaction of drift ETG modes and ZFs and show that ZFs
can be readily excited by the modulational instability.

This paper is organized as follows. In Sec.
\ref{SectionModelEquations} a set of normalized fluid equations
describing ETG drift modes -- ZF interaction is introduced. In
Sec. \ref{SectionModulationInstability}, the nonlinear dispersion
relation is obtained and the modulational instability growth rate
behavior is analyzed. Sec. \ref{SectionConclusions} contains
summary and conclusions.

\section{Model equations}
 \label{SectionModelEquations}
%
Assuming a slab two-dimensional geometry, charge quasineutrality
and the adiabatic ion responce, we consider the following
simplified model describing curvature driven ETG turbulence in the
inviscid limit (see, e.g. \cite{Gurcan1}):
\begin{eqnarray}
 \label{ModelEQ1}
  \frac{\partial}{\partial t}\left(1-\Delta_\perp\right)\varphi +
   \frac{\partial}{\partial y}\left(\varphi+P\right) -
   \left\{\varphi, \Delta_\perp\varphi\right\} = 0, \\
 \label{ModelEQ2}
  \frac{\partial}{\partial t}P - r\frac{\partial\varphi}{\partial
  y} - \left\{P, \varphi\right\} = 0,
\end{eqnarray}
where $\varphi$ and $P$ are the normalized electrostatic potential
and plasma pressure respectively, and $\{A, B\} =
\partial_xA\partial_yB - \partial_yA\partial_xB$ is the Jakobian.
In equations (\ref{ModelEQ1}) and (\ref{ModelEQ2}), following the
notations of Ref. \cite{Gurcan1}, we have rescaled the variables
as follows
\begin{displaymath}
    r = \frac{\epsilon_B\epsilon_{*e}}{\epsilon_{*i}^2},
\end{displaymath}
\begin{displaymath}
 \varphi = \frac{1}{\epsilon_{*i}}\frac{e\phi}{T_i}, \qquad P =
 \frac{\epsilon_B}{\epsilon_{*i}^2}\frac{P}{P_{i0}},
\end{displaymath}
\begin{displaymath}
 x = \frac{x'}{\rho_{s}\sqrt{\tau}}, \qquad
 y = \frac{y'}{\rho_{s}\sqrt{\tau}}, \qquad
 t = \epsilon_{*i}\omega_{Bi} t',
\end{displaymath}
$x'$, $y'$ and $t'$ being the original physical coordinates (with
$x'$ the poloidal and $y'$ the radial coordinate),
\begin{displaymath}
 \epsilon_{*i} = \frac{\rho_s \sqrt{\tau}}{L_n}, \qquad
 \epsilon_B = \frac{\rho_s \sqrt{\tau}}{L_B}, \qquad
 \epsilon_{*e} = \frac{\rho_s \sqrt{\tau}}{L_p},
\end{displaymath}
where $L_n$, $L_B$ and $L_p$ are the background gradient scales
for the density, magnetic field and pressure respectively,
$\rho_s$ is the ion gyroradius calculated at the electron
temperature $T_{e}$, and $\tau = T_i/T_e$.

In the linear limit, equations (\ref{ModelEQ1}) and
(\ref{ModelEQ2}) give the dispersion relation for ETG modes
\begin{equation}
  \label{LinearDispersion}
    \omega_{1,2} = \frac{k_y}{2\left(k^2+1\right)}
     \left[1\pm\sqrt{1-4r\left(k^2+1\right)}\right],
\end{equation}
where the plus sign describes the drift waves dispersion, while
the minus sign corresponds to convective cells. The linear
stability condition then reads as
\begin{displaymath}
  1 - 4r\left(k^2+1\right)\ge0.
\end{displaymath}
If $r>1/4$, the linear stability condition never holds, and ETG
mode is always unstable. For $r<1/4$, the stable ETG modes are
confined inside the region $k^2\le1/(4r) - 1$. Below we restrict
our analysis to the case of linearly stable ETG drift waves.

\section{Modulational instability of ETG drift waves and zonal flow generation}
 \label{SectionModulationInstability}
Assuming that the zonal flow varies on much larger timescale than
ETG drift waves do, the standard decomposition into fast
($\widetilde{...}$, ETG drift wave related) and slow
($\widehat{...}$, zonal flow related) motions can be performed.
The perturbations of electrostatic potential $\varphi$ and plasma
pressure $P$ are presented as a sum of fast and slow parts
\begin{displaymath}
    \varphi = \widehat{\varphi} + \widetilde{\varphi}, \qquad
    P = \widehat{P} + \widetilde{P}.
\end{displaymath}
Following the standard averaging procedure (recall that we
consider the region below the marginal stability boundary, i.e.
the subcritical turbulence) one gets the following set of
equations
\begin{eqnarray}
 \label{DecomposedSystemEQ1}
   \frac{\partial}{\partial t}\left(
   1-\Delta_\perp\right)\widehat{\varphi} = \left\{
   \widetilde{\varphi}, \Delta_\perp\widetilde{\varphi}\right\},\\
 \label{DecomposedSystemEQ2}
   \frac{\partial}{\partial t}\left(
   1-\Delta_\perp\right)\widetilde{\varphi} + \frac{\partial}{\partial
   y}\left( \widetilde{\varphi} + \widetilde{P}\right) =
   \left\{\widehat{\varphi},
   \Delta_\perp\widetilde{\varphi}\right\} + \left\{
   \widetilde{\varphi}, \Delta_\perp\widehat{\varphi}\right\},\\
 \label{DecomposedSystemEQ3}
   \frac{\partial}{\partial t} \widehat{P} = \left\{
   \widetilde{P}, \widetilde{\varphi}\right\}, \\
 \label{DecomposedSystemEQ4}
   \frac{\partial}{\partial t}\widetilde{P} - r\frac{\partial \widetilde{\varphi}}{\partial
   y}=\left\{\widehat{P},
   \widetilde{\varphi}\right\} + \left\{
   \widetilde{P}, \widehat{\varphi}\right\}.
\end{eqnarray}
When obtaining the system (\ref{DecomposedSystemEQ1}) -
(\ref{DecomposedSystemEQ4}), we have assumed that the mean field
is described by the same model equations as the fluctuations, but
is driven by the Reynolds stress. This follows from the assumption
of adiabatic ion response for the flow as well as fluctuations,
which is valid for the ETG modes. We describe the interaction
between drft ETG modes and ZF in terms of a four-wave coupling
scheme, i.e., each fluctuation is taken to be coherent. Then, the
ETG drift waves are considered as a superposition of the pump wave
$(\vec{k},\omega_k)$ and two sidebands $(\vec{k}_\pm,\omega_\pm)$,
i.e.,
\begin{equation}
 \label{DecompositionStarts}
   \widetilde{\varphi} = \widetilde{\varphi}_0 + \widetilde{\varphi}_+
 + \widetilde{\varphi}_-, \qquad
 \widetilde{P} = \widetilde{P}_0 + \widetilde{P}_+
 + \widetilde{P}_-,
\end{equation}
where
\begin{eqnarray}
  \widetilde{\varphi}_0 =
 \varphi_0\exp\left(\textrm{i}\vec{k}\vec{r}-\textrm{i}\omega_k
   t\right) + \textrm{c.c.}, \\
 \widetilde{\varphi}_\pm =
 \varphi_\pm\exp\left(\textrm{i}\vec{k_\pm}\vec{r}-\textrm{i}\omega_\pm
   t\right) + \textrm{c.c.}, \\
 \widetilde{P}_0 =
 P_0\exp\left(\textrm{i}\vec{k}\vec{r}-\textrm{i}\omega_k
   t\right) + \textrm{c.c.}, \\
 \widetilde{P}_\pm =
 P_\pm\exp\left(\textrm{i}\vec{k_\pm}\vec{r}-\textrm{i}\omega_\pm
   t\right) + \textrm{c.c.}
\end{eqnarray}
Zonal flow related electrostatic potential and pressure are taken
in the form
\begin{eqnarray}
\label{Pflow1}
 \widehat{\varphi} =
   \varphi_q\exp\left(\textrm{i}\vec{q}\vec{r}-\textrm{i}\Omega
   t\right) + \textrm{c.c.}, \\
\label{Pflow2}
 \widehat{P} =
   P_q\exp\left(\textrm{i}\vec{q}\vec{r}-\textrm{i}\Omega
   t\right) + \textrm{c.c.},
\end{eqnarray}
where $\vec{q} = (q, 0)$ is the wave vector of the zonal flow, and
the resonant conditions $\vec{k}_\pm = \vec{k} \pm \vec{q}$ and
$\omega_\pm = \omega_k \pm \Omega$ hold, where $\omega_k$ is the
ETG drift mode frequency, given by equation
(\ref{LinearDispersion}) with the $"+"$ sign in front of the
square root.

Substitution of (\ref{DecompositionStarts}) - (\ref{Pflow2}) into
the system (\ref{DecomposedSystemEQ1}) -
(\ref{DecomposedSystemEQ4}) gives:
\begin{eqnarray}
 \label{A1}
   \Omega\varphi_q = -\textrm{i}\frac{\left[\vec{k},
   \vec{q}\right]_z}{\left(q^2+1\right)}
     \left\{
       \left(k_+^2 - k^2\right)\varphi_0^*\varphi_+ -
       \left(k_-^2-k^2\right)\varphi_0\varphi_-^*
     \right\}, \\
 \label{A2}
   \varphi_+\left\{\omega_+\left(k_+^2+1\right)-k_{+y}\right\} -
   k_{+y} P_+ = \textrm{i}\varphi_0\varphi_q
   qk_y\left(k^2-q^2\right), \\
 \label{A3}
   \varphi_-^*\left\{\omega_-\left(k_-^2+1\right)-k_{-y}\right\} -
   k_{-y} P_-^* = \textrm{i}\varphi_0^*\varphi_q
   qk_y\left(k^2-q^2\right), \\
 \label{A4}
   \Omega P_q = \textrm{i}\left[\vec{k},
   \vec{q}\right]_z \left\{ \varphi_0 P_-^* + \varphi_+ P_0^* -
   \varphi_-^* P_0 - \varphi_0^* P_+ \right\}, \\
 \label{A5}
   \omega_+P_+ + k_{+y}r\varphi_+ =
   \textrm{i}qk_y\left[\varphi_qP_0 - \varphi_0P_q\right], \\
 \label{A6}
   \omega_-P_-^* + k_{-y}r\varphi_-^* = \textrm{i}qk_y \left[
   \varphi_qP_0^* - \varphi_0^*P_q\right].
\end{eqnarray}
Combining equations (\ref{A1})-(\ref{A6}), one can calculate the
amplitudes of the up-shifted and down-shifted satellites
\begin{equation}
 \label{PhiPlus}
   \varphi_{+} = \frac{\textrm{i}qk_y}{S_+}\left\{
   \varphi_q\varphi_0\left(k^2-q^2\right) +
   \frac{k_{+y}}{\omega_+}\left(\varphi_qP_0 -
   \varphi_0P_q\right)\right\},
\end{equation}
\begin{equation}
 \label{PhiMinusConj}
   \varphi_{-}^* = \frac{\textrm{i}qk_y}{S_-}\left\{
   \varphi_q\varphi_0^*\left(k^2-q^2\right) +
   \frac{k_{-y}}{\omega_-}\left(\varphi_qP_0^* -
   \varphi_0^*P_q\right)\right\},
\end{equation}
\begin{eqnarray}
 \label{PPlus}
   P_+ = \frac{\textrm{i}qk_y}{\omega_+}\left\{\varphi_qP_0 -
   \varphi_0P_q -
   \frac{k_{+y}r}{S_+}\varphi_q\varphi_0\left(k^2-q^2\right)
   \right. \\
 \nonumber
 \left. - \frac{k_{+y}^2r}{\omega_+S_+} \left(\varphi_qP_0 -
   \varphi_0P_q\right) \right\},
\end{eqnarray}
\begin{eqnarray}
 \label{PMinusConj}
   P_-^* = \frac{\textrm{i}qk_y}{\omega_-}\left\{\varphi_qP_0^* -
   \varphi_0^*P_q -
   \frac{k_{-y}r}{S_-}\varphi_q\varphi_0^*\left(k^2-q^2\right)
   \right. \\
 \nonumber
   \left. - \frac{k_{-y}^2r}{\omega_-S_-} \left(\varphi_qP_0^* -
   \varphi_0^*P_q\right) \right\},
\end{eqnarray}
where we have introduced the notation
\begin{displaymath}
   S_\pm = \omega_\pm\left(k_\pm^2+1\right) - k_{\pm}
   + \frac{k_{\pm}^2 r}{\omega_\pm}.
\end{displaymath}
Using the relation $P_0=-(rk_y / \omega_k)\,\varphi_0$ and
eliminating $P_{0}$, from equations (\ref{A1}), (\ref{A4}) and
(\ref{PhiPlus}) -- (\ref{PMinusConj}) one can get the nonlinear
dispersion relation in the form
\begin{equation}
 \label{a11a22_a12a21}
   a_{11}a_{22} - a_{12}a_{21} = 0,
\end{equation}
where
\begin{eqnarray}
 \label{a11new} \nonumber
     a_{11} = \Omega + \frac{\left[\vec{k},
     \vec{q}\right]_z^2}{q^2+1} \varphi_0^2 \left\{
     \left(k^2-q^2\right)\left(\frac{k_+^2-k^2}{S_+} -
     \frac{k_-^2-k^2}{S_-} \right) \right. \\
 \nonumber
   \left. - \frac{rk_y^2}{\omega_k}\left(
     \frac{k_+^2 - k^2}{\omega_+S_+} - \frac{k_-^2 -
     k^2}{\omega_-S_-} \right)
   \right\},
\end{eqnarray}
\begin{eqnarray}
 \label{a12new} \nonumber
   a_{12} = -\frac{\left[\vec{k},
     \vec{q}\right]_z^2}{q^2+1}k_y\varphi_0^2\left\{
     \frac{k_+^2-k^2}{\omega_+S_+} - \frac{k_-^2-k^2}{\omega_-S_-}
     \right\},
\end{eqnarray}
\begin{eqnarray}
 \label{a21new} \nonumber
   a_{21} = \frac{rk_y}{\omega_k}\left[\vec{k},\vec{q}\right]_z^2 \varphi_0^2 \left\{
   \left(\omega_k\left(k^2-q^2\right) +
   \frac{rk_y^2}{\omega_k}\right)\left(\frac{1}{\omega_+S_+} -
   \frac{1}{\omega_-S_-}\right) \right. \\
 \nonumber
   \left. -
   \left(k^2-q^2\right)\left(\frac{1}{S_+} - \frac{1}{S_-}\right)
   + \frac{1}{\omega_+}\left(1-\frac{rk_y^2}{\omega_+S_+}\right) -
   \frac{1}{\omega_-}\left(1-\frac{rk_y^2}{\omega_-S_-}\right)
   \right\},
\end{eqnarray}
\begin{eqnarray}
 \label{a22new} \nonumber
   a_{22} = -\Omega + \left[\vec{k},\vec{q}\right]_z^2 \varphi_0^2
   \left\{ \frac{rk_y^2}{\omega_k}\left(\frac{1}{\omega_+S_+} -
   \frac{1}{\omega_-S_-}\right) \right. \\
 \nonumber
 \left.
   + \frac{1}{\omega_+}\left(1-\frac{rk_y^2}{\omega_+S_+}\right) -
     \frac{1}{\omega_-}\left(1-\frac{rk_y^2}{\omega_-S_-}\right)
   \right\}.
\end{eqnarray}
Generally, ZF dispersion which follows from the equation
(\ref{a11a22_a12a21}) is a fourth-order in $\Omega$, which, in
principle, allow one to treat it analytically. We present below an
analysis of nonlinear dispersion relation for the case when the
ETG drift mode does not have the $k$-component in the direction of
inhomogeneity, i.e. for $k_x = 0$. We also treated the case $k_x
\ne 0$ and found that inclusion of nonzero $k_x$ does not
introduce any qualitative change to the growth rate behavior
except for the decrease of it's magnitude. Besides that, it turns
out that in all studied cases the pump wave with $k_x=0$ ensures
the largest growth rate in the explored region of linear
stability.

\begin{figure}
  \includegraphics{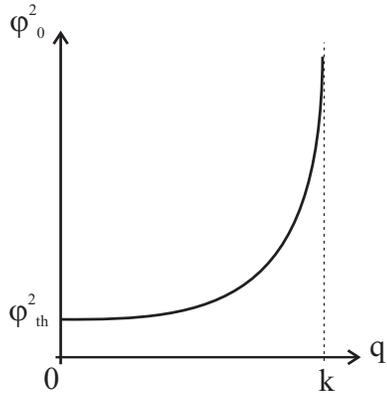}
  \caption{Schematic plot of the modulational instability threshold
           dependence on the scale of the excited zonal flow.
           $\varphi_{th}^2$ is given by (\ref{Threshold}).}\label{FigureThreshold}
\end{figure}

For the case $k_x=0$, 
the nonlinear dispersion relation is reduced to the biquadratic
equation
\begin{equation}
  \label{BiquadraticEquation}
    c_4\Omega^4 + c_2\Omega^2 + c_0 = 0,
\end{equation}
where
\begin{eqnarray}
 \nonumber
   c_0 =
   4k^2q^6\left(k^2-q^2\right)
   \left(k^2+q^2+1 \right)\varphi_0^4 \\
 \nonumber
   + 2k^2q^4\left[k^2-2k\left(1+2k^2\right)\omega_k + \left(
   1+5k^2+4k^4-2\left(q^2+q^4\right)\right)
   \omega_k^2\right]\varphi_0^2 \\
 \nonumber
  -q^4\left(1+q^2\right)\omega_k^4, \\
\nonumber
   c_2 = 2 k^2 q^2 \left(k^2+q^2+1\right)
   \left(k^2+2 \left(q^4+q^2\right)+1\right) \varphi_0^2 \\
 \nonumber
   + \left(q^2+1\right) \left[k^2-4 k \omega_k
   \left(k^2+q^2+1\right) \right.\\
 \nonumber
   \left. +2 \left(k^2+q^2+1\right) \left(2 \left(k^2+1\right)+q^2\right) \omega_k
   ^2\right], \\
 \nonumber
   c_4 = -\left(q^2+1\right) \left(k^2+q^2+1\right)^2,
\end{eqnarray}
which yields the modulational instability growth rate
\begin{equation}
  \label{GrRateExplicit}
    \gamma = \max \,\,\mathrm{Im} \left(
       \pm \sqrt{\frac{-c_2 \pm \sqrt{c_2^2-4c_0c_4} }{2c_4}}
      \right).
\end{equation}
The analysis of the growth rate (\ref{GrRateExplicit}) shows that
the modulational instability can be excited ($\gamma>0$) only
above some threshold in the input power $\varphi_0^2$. This
threshold is a function of the scales $k$ and $q$ of ETG drift
wave and zonal flow. The minimum threshold corresponds to the case
$q\rightarrow0$ and is given by
\begin{equation}
  \label{Threshold}
     \varphi_{th}^2=\frac{v_{ph} ^4}{2 \left(1-2(1+2 k^2)v_{ph}
     +(1+k^2)(1+4k^2)v_{ph}^2\right)},
\end{equation}
where $v_{ph}=\omega_k/k$ is the phase velocity of ETG drift wave.
As the zonal flow wavenumber $q$ approaches $k$, the instability
threshold goes to infinity, as it is illustrated in Figure~
\ref{FigureThreshold}.

\begin{figure}
  \includegraphics{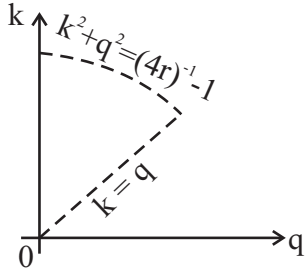}
  \caption{Modulational instability domain  in the $(k, q)$-plane
   is confined by the dashed lines.}  \label{FigureDomain}
\end{figure}

Present analysis is restricted to the case when the pump ETG drift
wave, as well as both sidebands are linearly stable, i.e. for each
value of $r$ parameter, all considered scales $k$ and $q$ are
confined in the $(k,q)$ - plane inside the circle
\begin{equation}
 \label{TreatedRegion}
   k^2 + q^2 = \frac{1}{4r} - 1.
\end{equation}
Thus, the modulational instability parameter range under
consideration is limited. Instability domain in $(k, q)$-plane is
outlined in Figure ~\ref{FigureDomain}. It is clear that circle
shrinks to the origin when $r$ approaches $1/4$.

\begin{figure}
  \includegraphics{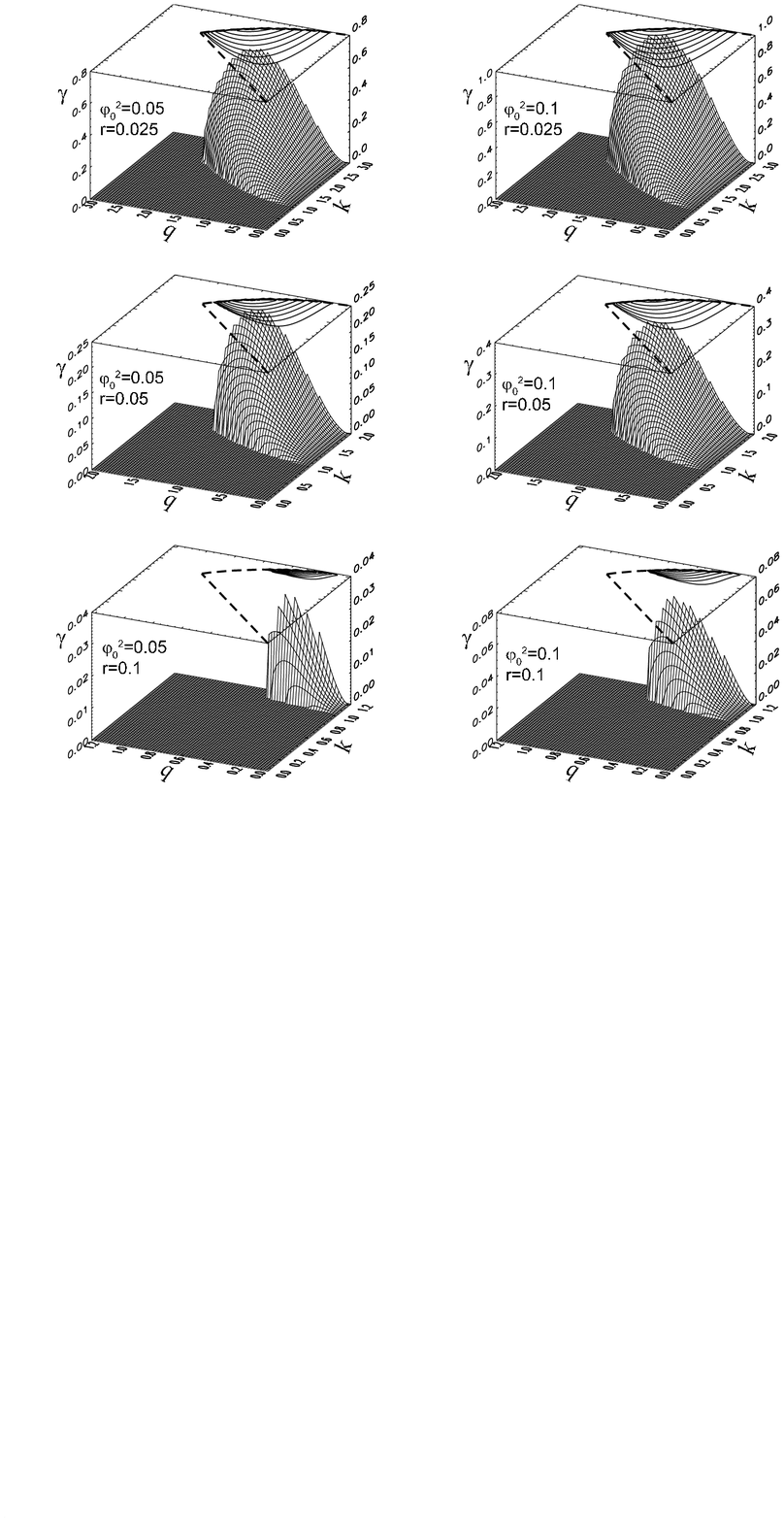}
  \caption{ Modulational instability growth rate $\gamma$ given by (\ref{GrRateExplicit})
  as a function of $(k, q)$ for different $\varphi_0^2$ and $r$.
  Dashed lines in the contour plane indicate the instability domain.} \label{GrRatesPlate}
\end{figure}

In the Figure~\ref{GrRatesPlate}, modulational instability growth
rates are plotted as a functions of $k$ and $q$ scales of ETG
drift wave and zonal flow respectively for different values of the
parameter $r$ and pump wave amplitude $\varphi_0^2$. Dashed lines
in the contour planes confine the considered instability domain
(schematically outlined in Figure~\ref{FigureDomain}). Outside the
circle given by the equation (\ref{TreatedRegion}), the growth
rate was artificially forced to zero due to linear stability
requirements. It is seen that as the parameter $r$ grows, the
region of instability contracts and the growth rate magnitude goes
down. In all studied cases, the maximum growth rate always
corresponded to the $(k, q)$ - pair from the circle
(\ref{TreatedRegion}) so that for the fixed $k$ there is an
optimal (i.e. corresponding to the maximum growth rate) scale $q$
of the generated ZFs. With the increase of a pump wave amplitude,
the growth rate as well as the instability domain increases,
however in the region of physically appropriate values of
$\varphi_0^2$, no qualitative changes to the growth rate behavior
were found.

\section{Summary and conclusions}
  \label{SectionConclusions}

In the present paper, we have considered the nonlinear interaction
between drift waves and ZFs in subcritical ETG turbulence. We have
derived a set of equations describing the dynamics of nonlinearly
coupled ETG modes and ZFs. Assuming that the spectrum of
fluctuations is sufficiently narrow, we analyzed the obtained set
of equations in terms of a four-wave coupling scheme and obtained
the nonlinear dispersion relation. We have shown that this
dispersion relation predicts the modulational instability of a
finite amplitude monochromatic drift wave. Thus, ETG drift
fluctuations can be destabilized by the four-wave interaction
mechanism with simultaneous generation of ZFs. We have found the
threshold of the modulational instability and the dependence of
the instability growth rate on spatial scales of ETG drift waves
and excited ZFs. For the fixed wavenumber of the ETG mode $k$, the
growth rate always has a maximum which is achieved for some
intermediate value of ZF wavenumber $q$. When the amplitude of the
ETG pump wave increases (as well as when the parameter $r$
decreases), the region of modulational instability widens towards
small $k$-scales. The present results thus demonstrate that ZFs in
subcritical ETG turbulence can be excited by the modulational
instability.


\end{document}